\documentstyle[prb,twocolumn,aps]{revtex}
\input{psfig}

\preprint{}
\begin{document}
\draft
\title{Superconducting $d$-wave junctions: The
disappearance of the odd ac components}
\author{Tomas L\"ofwander,$^a$ G\"oran Johansson,$^a$ Magnus Hurd,$^{a,b}$ 
and G\"oran Wendin$^a$}
\address{
$^a$ Division of Microelectronics and Nanoscience, Department of Physics,\\
Chalmers University of Technology and G\"oteborg University,\\
S-412 96 G\"oteborg, Sweden\\
$^b$ Department of Technology and Natural Science, 
Halmstad University, S-301 18 Halmstad, Sweden}
\date{30 September 1997}
\maketitle
\begin{abstract}

We study voltage-biased superconducting planar $d$-wave junctions 
for arbitrary transmission and arbitrary orientation of the order 
parameters of the superconductors.
For a certain orientation of the superconductors the odd ac components
disappear, resulting in a doubling of the Josephson frequency.
We study the sensitivity of this disappearance to orientation and
compare with experiments on grain boundary junctions.
We also discuss the possibility of a current flow parallel to the junction.
\end{abstract}

 
\narrowtext

Some of the hole-doped high-$T_c$ cuprates show features that cannot be 
explained using an $s$-wave order parameter.\cite{Harl,Tsue} Instead 
various phase-sensitive experiments involving Josephson junctions suggest 
that the order parameter changes sign at certain regions of the Fermi 
surface. The outcome of these and other experiments makes $d$-wave 
symmetry a particularly interesting candidate for the symmetry of the 
superconducting order parameter in the high-$T_c$ cuprates.~\cite{Harl}

Hu's prediction of midgap states (MGS) at interfaces/surfaces of $d$-wave 
superconductors\cite{Hu} called for a reinvestigation of a number of 
transport problems involving $d$-wave superconductors. This reinvestigation 
has to some extent been carried out, starting with the normal 
metal-superconductor (NS) junction,\cite{TanKas1,XMT} followed by the 
dc Josephson effect\cite{BBR,TanKas3,SamDat} and the ac Josephson 
effect.\cite{BarSvi,Hurd} It has been found that the zero-energy
MGS influence the current-voltage relation (the NS-case and the ac Josephson 
effect), the temperature dependence, and the current-phase relation 
(the dc Josephson effect).

The origin of MGS is due to normal scattering of the quasiparticles at
interfaces/surfaces. Since the quasiparticle changes its momentum when
scattered, it will in the $d$-wave case experience a different order
parameter after scattering. When there is a sign change of the order
parameter after scattering, a bound state is formed at the interface/surface
of the superconductor.\cite{Hu}

\begin{figure}
\centerline{\psfig{figure=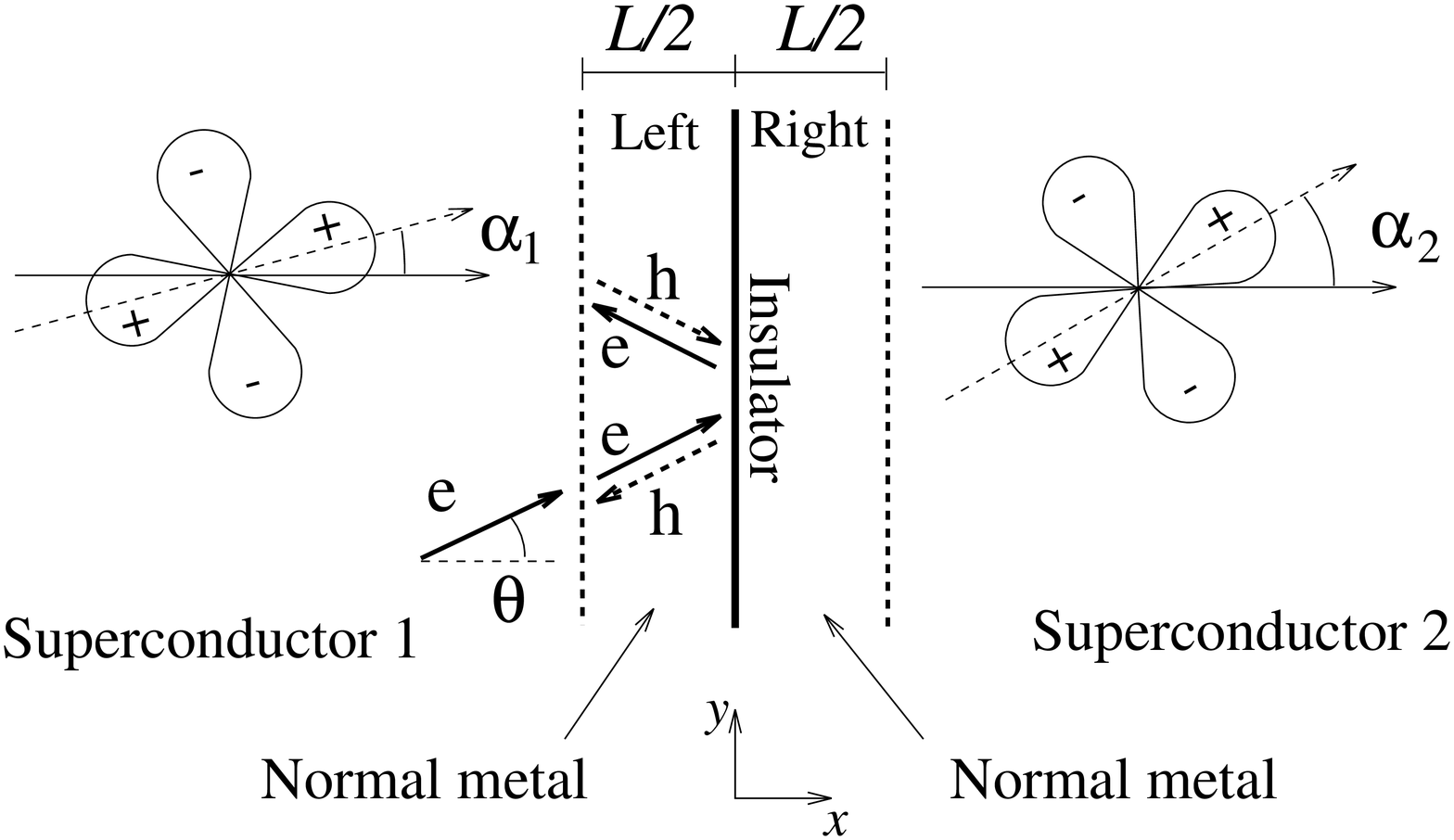,width=8cm}}
\caption{Layout of the junction. The superconductors can be rotated to 
an angle $\alpha_{1,2}$. The strength of the barrier can be tuned all 
the way from the ballistic to the insulating case.}
\label{fig1}
\end{figure}

In this paper we investigate the ac Josephson effect in planar $d$-wave 
junctions, calculating the first Fourier components of the current.
The order parameter $\Delta(\theta)$ for an anisotropic superconductor 
depends on the angle $\theta$ of incidence for the quasiparticle 
approaching the junction. For the $d$-wave case the order parameter is 
taken to be $\Delta(\theta)=\Delta_0\cos[2(\theta-\alpha)]$, where 
$\alpha$ is the angle of orientation of the superconducting order 
parameter with respect to the  junction interface as explained in 
Fig.~\ref{fig1}. The cases $\alpha=0$ and $\alpha=\pi/4$ correspond 
to $d_{x^2-y^2}$ and $d_{xy}$ symmetry, respectively. For the $s$-wave 
case, $\Delta(\theta)=\Delta_s$.

For conventional superconducting junctions with $s$-wave electrodes 
the frequency of the non-stationary Josephson current is 
$\omega_J=2eV/\hbar$.  This means that the Josephson current 
perpendicular to the junction  can be decomposed into components 
with frequency parts $m \omega_J$ ($m$ is an integer):\cite{BSW,AveBar}
\begin{equation}
I(t)=\sum_m I_m e^{i m \omega_J t}.
\label{FourCurr}
\end{equation}
The zero-frequency component has been studied recently for the $d$-wave 
case.\cite{Hurd} In this paper we extend the analysis to include the
nonzero frequency parts of the current.

In the $s$-wave case one finds that all components $I_m$ in 
Eq. (\ref{FourCurr}) are nonzero although they decrease with 
increasing $m$. This is not necessarily the case when $d$-wave 
superconductors are introduced in the junction. Consider the $d$-wave
junction and let the order parameter of the left (right) superconductor 
be $\Delta_{1}(\theta)$ [$\Delta_2(\theta)$]. For orientations of the 
superconductors where $\Delta_1(\theta)=\Delta_1(-\theta)$ 
(the case for $s$-wave and $d_{x^2-y^2}$ orientation of the left 
superconductor) and $\Delta_2(\theta)=-\Delta_2(-\theta)$ 
($d_{xy}$ orientation of the right superconductor) we find that 
$I_m=0$ for odd $m$ and therefore the Josephson frequency is doubled. 
This doubling of the Josephson frequency has previously been inferred 
from the stationary Josephson effect,\cite{Zago} where one can show 
that in the situation described above the Josephson current is 
$\pi$-periodic in the phase rather than $2\pi$-periodic. This change 
of period in the phase would then change the Josephson frequency accordingly. 

Considering the nonstationary Josephson effect, we may calculate the 
Fourier components and explore the sensitivity of the odd ac components 
to deviation from orientations where disappearance of the odd ac components 
occur.  This is the main subject of this paper and is not possible to treat 
within the framework of the stationary Josephson effect.

In passing, we note that the disappearance of the odd ac
components is not related to the presence of MGS itself; even for the 
ballistic junction (when there are no MGS present) the odd ac components 
disappear in the situation described above.

In our calculation self-consistency of the superconducting order parameter 
is not fulfilled. Recent investigations in the tunneling limit have 
considered self-consistency and found that interface states with nonzero 
energy appear along with the zero-energy MGS.\cite{BarSvi} In this paper 
we neglect these effects. 

The system that we consider is shown in Fig.~\ref{fig1}. Two
superconductors are separated by a barrier represented by a
scattering matrix, which for electrons is
\begin{equation}
S_e(\theta)=\left(
\begin{array}{cc}
r(\theta) & t(\theta)\\
t(\theta) & -r^*(\theta)t(\theta)/t^*(\theta)
\end{array}
\right).
\end{equation}
The scattering matrix for holes is $S_h(\theta)=S_e^*(\theta)$.
The normal reflection at the barrier changes the angle from $\theta$ to
$\bar{\theta}=\pi-\theta$.
All barred quantities in this paper are related to $\bar{\theta}$.

We solve the time-dependent Bogoliubov-de Gennes equation for 
$d$-wave superconductors piecewise in each region. 
The details of the method can be found in e. g. Ref.~\onlinecite{Hurd}. 
The order parameter is assumed to be
\begin{equation}
\Delta({\bf k},{\bf x},t)=
\left\{ \begin{array}{ll}
\Delta_1(\theta)\;, & x<-L/2 \\
0\;, & |x|<L/2 \\
\Delta_2(\theta)e^{i(\phi_0+2eVt/\hbar)}\;, & x>L/2.
\end{array}
\right.
\label{ppot}
\end{equation}
The overall phase is not important
and we therefore choose $\Delta_{1}$ real and let the phase difference
between the superconductors be $\phi_0$. In Eq.~(\ref{ppot}) it is
understood that we only consider ${\bf k}$-vectors with $|{\bf k}|=k_F$.

Solving the scattering problem, we inject an electron-like quasiparticle 
from the left superconductor at energy $E$ and angle $\theta$. Since there 
is a voltage bias over the junction, the quasiparticle goes through 
multiple Andreev reflections (MAR). The barrier is modeled as a 
$\delta$-function potential of strength $H$ with transmission amplitude 
$t(\theta)=i\cos\theta/(i\cos\theta-Z)$ and reflection amplitude 
$r(\theta)=Z/(i\cos\theta-Z)$, where $Z$ is defined as $Z=2mH/\hbar^2$.
Our calculations include normal regions to the left and to the right of
the barrier, but the length $L$ is put to zero at the end of the calculation. 
Therefore, our method follows closely the one presented by Averin and 
Bardas.~\cite{AveBar} However, the same results would have been obtained 
if we had used the method of Bratus, Shumeiko, and Wendin.~\cite{BSW,SBW}

It is convenient to calculate the current in the normal region to the
left of the barrier. The wave function in this region is
\begin{equation}
{\Psi}^{\rightarrow} =
\sum_{n} 
\left\lgroup
\begin{array}{c} a_{2n}^{\rightarrow}e^{i {\bf k}\cdot{\bf x}}
+ d_{2n}^{\rightarrow} e^{i {\bf\bar{k}}\cdot{\bf x}} 
\\
b_{2n}^{\rightarrow} e^{i {\bf k}\cdot{\bf x}} 
+c_{2n}^{\rightarrow} e^{i {\bf \bar{k}}\cdot{\bf x}} 
\end{array}
\right\rgroup
e^{-i (\frac{E_{2n} t}{\hbar}+n\phi_0)} \; ,
\label{wfnL}
\end{equation}
where the momentum is defined by 
${\bf k}=(k_x,k_y)=k(\cos\theta,\sin\theta)$ and
$\bar{\bf k}=(-k_x,k_y)=k(\cos\bar{\theta},\sin\bar{\theta})$. We have 
introduced the energy $E_n=E+neV$, where $n$ is an integer. Note that we 
have separated out the phase difference $\phi_0$ between the superconductors, 
making the coefficients $a$, $b$, $c$, and $d$ independent of $\phi_0$. 
Matching the wave functions of the different regions one gets the 
coefficients $a$, $b$, $c$, and $d$ as written down in Ref.~\onlinecite{Hurd}.

The next step is to calculate the current by inserting the wave function
into the following formula for the current density
\begin{equation}
j=\frac{e\hbar}{m}\sum_n \tanh\left(\frac{-E_n}{2k_B T}\right)\mbox{Im}\left\{
u_n^*\nabla u_n+v_n^*\nabla v_n \right\},
\label{currentdensity}
\end{equation}
where $(u_n,v_n)$ is the solution of the BdG equation.
The sum is over all incoming electron-like quasiparticles for both negative
and positive energies. There are different ways of writing the current 
density. Using the completeness relation for solutions of the BdG 
equation~\cite{Bardeen} one can show that Eq. (\ref{currentdensity}) is 
equivalent to the current formulas written down in previous 
work.\cite{BSW,HDB,BSBW}

Including also the left-moving quasiparticles we can determine the
expression for the ac components $I_m$ in Eq. (\ref{FourCurr}):
\begin{eqnarray}
&&\frac{I_m(V)}{\sigma_0}=C_m(V)
e^{i(m\phi_0-\alpha_{m})},\ \ m\neq0 \nonumber\\
&&C_m(V)=\sqrt{A_m^2(V)+B_m^2(V)},\ \ 
\alpha_m=\arctan\frac{B_m(V)}{A_m(V)},\nonumber\\
&&\sigma_0=L_y\frac{2^{5/2}em^{1/2}E_{F}^{1/2}\Delta_0D}{h^2},
\label{FourComponents}
\end{eqnarray}
where $D=\int d\theta |t|^2 \cos\theta/2$ is the transmission probability 
through the barrier averaged over all angles $\theta$, 
and $L_{y}$ is the junction length in the $y$-direction.~\cite{correction} 
In Eq. (\ref{FourComponents}) we introduced $A_m(V)$ and $B_m(V)$ 
which are the coefficients in the cosine and sine expansion
of the ac current. The coefficients are defined as
\begin{eqnarray}
A_{m}(V)=&&\frac{1}{4D}\int_{-\pi/2}^{\pi/2}d\theta \cos\theta\
\int_{-\infty}^{\infty}\frac{dE}{\Delta_0}
\tanh\left(\frac{-E}{2k_B T}\right)\nonumber\\
&&\times\sum_\tau \tau N_\tau (E){\mbox{Re}}\left\{
T^{\tau}(E,\theta,m)\right\},\nonumber\\
B_{m}(V)=&&-\frac{1}{4D}\int_{-\pi/2}^{\pi/2}d\theta \cos\theta
\int_{-\infty}^{\infty}\frac{dE}{\Delta_0}
\tanh\left(\frac{-E}{2k_B T}\right)\nonumber\\
&&\times\sum_\tau N_\tau (E){\mbox{Im}}\left\{
T^{\tau}(E,\theta,m)\right\},
\label{AmBm}
\end{eqnarray}
where $\tau=+$ $(-)$ means right (left) movers,
$N_{\tau}(E,\theta)$ is the bulk superconducting density of states 
evaluated in superconductor $1$ ($2$) for $\tau=+$ $(-)$, and
\begin{eqnarray}
T^\tau(E,\theta,m)=&&\sum_{n=-\infty}^{\infty}
(a_{2n+2m,\tau}^{*}a_{2n,\tau}-d_{2n+2m,\tau}^{*}d_{2n,\tau}\nonumber\\
&&+b_{2n+2m,\tau}^{*}b_{2n,\tau}-c_{2n+2m,\tau}^{*}c_{2n,\tau}).
\label{Teh}
\end{eqnarray}

The phase $m\phi_0-\alpha_{m}$ in Eq. (\ref{FourComponents})
defines the ($m$-dependent) time when the measurement is started, and
may not be easily observed in a transport experiment. To illustrate 
our point that $I_m$ (where $m$ is an odd number) disappears for certain 
orientations it is enough to plot the amplitude $C_m(V)$.

We now investigate the ac components. For $s$-wave superconductors we have
the ac Josephson effect where the Josephson frequency is $\omega_J$. In the
case of $d$-wave superconductors this may not be true,
as discussed above. For the case $d_{x^{2}-y^{2}}/d_{xy}$ we find $C_1=0$ 
and $C_2\neq0$, meaning that the Josephson frequency for this configuration 
is $2\omega_J$. Decreasing $\alpha_2$ from $\pi/4$ (the case $d_{xy}$), 
the magnitude  $C_m$ of the odd ac components increases from zero 
(and therefore  reestablishes the ordinary Josephson frequency $\omega_J$) 
until  $\alpha_2$ reaches $0$, in which case the situation more resembles 
the $s$-wave case. To understand this effect we study the case 
$d_{x^{2}-y^{2}}/d_{xy}$. The expression for the current involves an 
integration over both positive and negative injection angles $\theta$. 
We may explicitly calculate the current for positive and negative angles 
separately, and then add them to get the final result. When changing the 
angle $\theta$ to $-\theta$ we must change the phase $\phi_0$ between the 
superconductors by $\pi$, because 
$\Delta_2(-\theta)=-\Delta_2(\theta)=\Delta_2(\theta)e^{i\pi}$ (meaning
$\phi_0\rightarrow\phi_0+\pi$). The effect of this is seen in the wave 
function in Eq.~(\ref{wfnL}), where we get an extra phase $-in\pi$. 
This phase, separated out together with $\phi_0$, will pass through
the calculation and finally show up in the expression for the current:
\begin{eqnarray}
I(V,t)&=&\sum_m(I_m(V,\theta>0)+I_m(V,\theta<0))e^{im\omega_J t}\nonumber\\
&=&\sum_m I_m(V,\theta>0)(1+e^{im\pi})e^{im\omega_J t},
\label{current_symm}
\end{eqnarray}
where the amplitudes $I_m(V,\theta>0)$ [$I_{m}(V,\theta<0)$] now are defined
with an integration over positive (negative) angles only. The factor 
$(1+e^{im\pi})$ is zero for odd $m$ and two for even $m$. This implies 
a doubling of the Josephson frequency for the case $d_{x^{2}-y^{2}}/d_{xy}$.

\begin{figure}[t]
\centerline{\psfig{figure=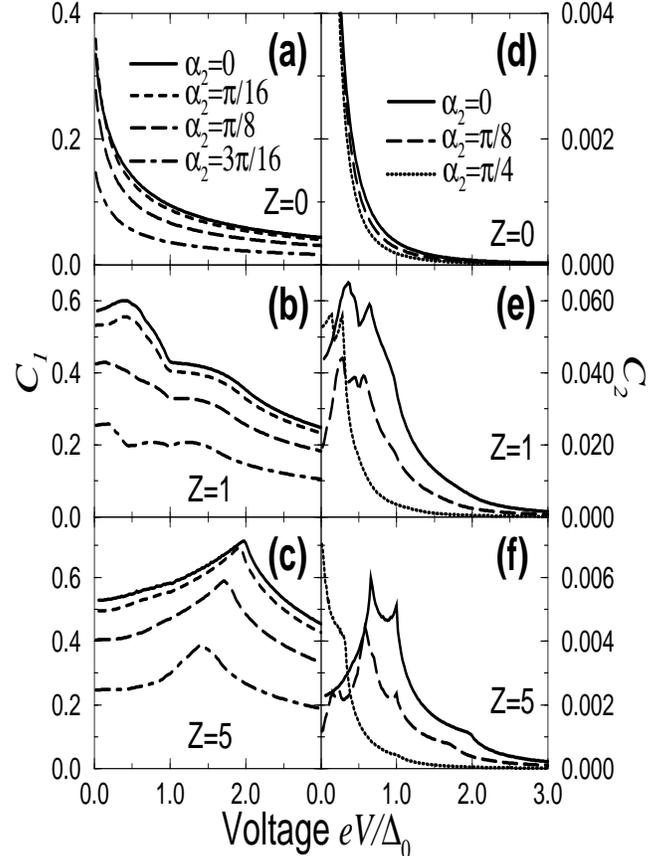,width=10cm,height=11cm}}
\caption{Here we show the first two Fourier components as
a function of voltage for three different barrier strengths $Z$.
The orientation of the left superconductor is fixed at 
$\alpha_1=0$, while we vary the orientation of the right 
superconductor from $\alpha_2=0$ to $\alpha_2=\pi/4$. 
The first component $C_1$ is identically zero for 
$\alpha_2=\pi/4$ as explained in the text. Zero temperature is assumed.}
\label{fig2}
\end{figure}

We present the ac components for different configurations of the 
superconductors and three different barrier strengths $Z$. The plots are shown
in Fig.~\ref{fig2}, and are only for $V>0$ since one can show that
$I(-V)=-I(V)$. For the $d_{x^{2}-y^{2}}/d_{xy}$ junction we only plot 
the first non-zero amplitude $C_{2}$. For the 
$d_{x^{2}-y^{2}}/d_{\alpha_2\neq\pi/4}$ junction we plot
the components $C_{1}$ and $C_{2}$ since they are both non-zero.

The results show that the amplitude $C_1$ decreases smoothly as the angle
$\alpha_2$ is increased from $0$ to $\pi/4$. There is  subharmonic gap 
structure (SGS) for the case of non-zero barrier strength at the same 
voltages as in the $s$-wave case [see Figs.~\ref{fig2}(b) and 
\ref{fig2}(c)].~\cite{HDB} Angular averaging partly 
washes out the SGS.~\cite{Hurd} There is no SGS for the ballistic case 
$Z=0$ at zero temperature [see Fig.~\ref{fig2}(a)].~\cite{HDB}

\begin{figure}
\centerline{\psfig{figure=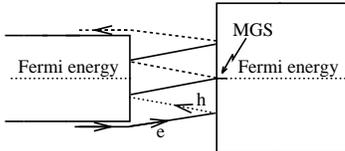,height=2.0cm}}
\caption{The trajectory above gives rise to a large amplitude of the
second component $C_2$ at small voltages, see Fig.~\ref{fig2}(f). 
There is a resonance when the trajectory starts from the gap 
edge of the left superconductor, hits MGS, and finally ends 
at the gap again.}
\label{fig3}
\end{figure}

Considering the second Fourier component $C_2$, we first note that this
component is much smaller than $C_1$ simply because $C_2$ corresponds
to a higher order term in the expansion of the current in Eq. (\ref{FourCurr}).
The curves depicted in Figs.~\ref{fig2}(e) and \ref{fig2}(f) again show
SGS [no SGS in the ballistic case $Z=0$ shown in Fig.~\ref{fig2}(d)].\cite{HDB}
When $\alpha_2$ is close to $\pi/4$ and $Z$ is large [see Fig.~\ref{fig2}(f)],
MGS clearly influence the curves of $C_2$, resulting in an increase in $C_2$
for small voltage. MGS appear as resonances in the scattering states of
Eq. (\ref{wfnL}) when the order parameter experienced by the quasiparticle 
changes sign after normal reflection at the barrier. 
Because of the large density of states at the gap, a 
quasiparticle trajectory passing through the MGS as the one
depicted in Fig.~\ref{fig3} will give rise to a large contribution 
to the current. The resonant trajectory originates from the gap edge
only when $3eV=\Delta_1=\Delta_0\cos 2\theta$ as seen in Fig.~\ref{fig3},
meaning that the upper limit for this particular process is $\Delta_0/3$.

Another quasiparticle trajectory passing through MGS is the one travelling
back and forth within the gaps twice instead of six times as shown in
Fig.~\ref{fig3}. This process will give rise to a smaller current
contribution to $C_2$ than the one shown in Fig.~\ref{fig3} for the
following reason. Calculating $C_2$ we study energies 
$E_{2n}$ and $E_{2n+4}$. For the process corresponding to 
$eV=\Delta_0 \cos 2\theta$, $E_{2n+4}$ will be above the gap region, 
and therefore this process will be subject to two additional non-perfect 
(out of gap) Andreev reflections which suppress the contribution.

The condition $eV=\Delta_1=\Delta_0\cos 2\theta$ for the process shown
in Fig.~\ref{fig3} does not depend on $\Delta_2$. This means that SGS
corresponding to this process is present for all angles $\alpha_2$
but the amplitude is smaller when $\alpha_2\neq\pi/4$ since the
resonant MGS do not occur at as many injection angles $\theta$ in this
case. For $\alpha_2=0$ the structure is absent since there is no MGS 
in this case. Larger transmission (smaller $Z$) broadens MGS, making 
the effects of MGS less drastic, as seen when comparing Fig.~\ref{fig2}(e) 
with Fig.~\ref{fig2}(f). The first component $C_1$ shows no increase in 
amplitude at small voltages. Actually, there is similar SGS but this is 
smaller in amplitude than the ordinary background and does not dominate 
in the same way as for $C_2$.

Considering the case of two $d_{xy}$ superconductors, we get a phase shift 
of $\pi$ in both superconductors when we let $\theta\rightarrow-\theta$, 
leading to  no overall phase difference. Therefore the first ac component 
$C_1$ does not vanish in this case.

In experiments on high-$T_c$ $d_{x^2-y^2}/d_{xy}$ grain boundary
junctions,\cite{Char} Shapiro steps have been found at voltages 
corresponding to a Josephson frequency of $2\omega_J$. As pointed 
out in Ref.~\onlinecite{Char} this effect may be explained by assuming 
that parallel junctions are formed at the grain boundary. According to 
our results it is possible that the doubled Josephson frequency is due 
to $d$-wave symmetry of the order parameter.

Very recently, it has been shown that there is a nonzero ground state 
current parallel to the junction~\cite{Sigrist} in the stationary case 
(zero voltage) for a $s/d_{xy}$ junction. Inspired by these investigations, 
we have used our formalism to calculate the parallel current for the 
non-stationary case. For junctions with two $d$-wave superconductors 
we find both time-independent ($m=0$) and time-dependent ($m\neq 0$) 
currents in the general case. For the $y$-direction current the even 
components disappear in the $d_{x^2-y^2}/d_{xy}$ geometry while the 
odd components are non-zero. This happens because we get a minus sign 
between the two terms in Eq. (\ref{current_symm}) instead of the plus 
sign, making the $x$-direction and the $y$-direction currents 
complementary. This resembles the stationary case where the parallel 
current appears when the perpendicular current is zero.~\cite{Sigrist}

Also for the $N/d_{\alpha}$ junction we find a time independent ($m=0$; 
no ac in the $NS$ case) current density in the $y$-direction at $x=0$ 
if the up/down symmetry of the junction is broken by the order parameter, 
which is the case when the angle $\alpha$ has a value between $0$ and 
$\pi/4$. For $\alpha=0$ and $\alpha=\pi/4$ the current density in the 
$y$-direction is zero. The $N/d_\alpha$ case has recently been 
discussed.\cite{Sauls} One notes that even for an asymmetric $s$-wave 
order parameter there are current contributions in the $y$-direction 
for the $NS$ case.

In summary, we have calculated the ac components for voltage biased
$d$-wave junctions. We find that the odd ac components
of the current in the $x$-direction vanish if the orientations of
the $d$-wave superconductors is $\alpha_1=0$ and $\alpha_2=\pi/4$.
For this case the Josephson frequency is doubled, which has also been
found in experiments on grain boundary junctions.\cite{Char}
The reason for this effect is that the contributions from injection
angles $+\theta$ and $-\theta$ have exactly the same magnitudes, 
but opposite signs. The angle average will then cancel the odd components.
We have also discussed the possibility of current flow 
in the direction parallel to the junction.

It is a pleasure to thank Z. Ivanov and V. S. Shumeiko for many
useful discussions on this work. The work has been supported by the Swedish
Natural Science Research Council. One of the authors (GW) acknowledges
support from the European Union Science and Technology Grant Programme
in Japan and the hospitality of the NTT Basic Research Laboratories.


\begin{references}
\bibitem{correction} There should be an extra factor of two in the 
definition of $\sigma_0$ written down in Ref.~\onlinecite{Hurd}.
\end{references}
\end{document}